\documentclass[conference]{IEEEtran}

\usepackage{caption}
\usepackage{graphicx}
\usepackage{multirow}
\begin{document}

\title{Asynchronous Circuits as an Enabler of Scalable and Programmable Metasurfaces\vspace{-1\baselineskip}}

\author{\IEEEauthorblockN{Loukas Petrou, Petros Karousios and Julius Georgiou}
\IEEEauthorblockA{Dept. of Electrical and Computer Engineering, University of Cyprus\\
Email: \{lpetro02, pkarou01, julio\}@ucy.ac.cy}
}

\maketitle

\begin{abstract}

Metamaterials and metasurfaces have given possibilities for manipulating electromagnetic (EM) waves that in the past would have seemed impossible. The majority of metasurface designs are suitable for a particular frequency and angle of incidence. One long-sought objective is the design of programmable metasurfaces to dynamically manipulate a variety of incoming EM frequencies and angles. In order to do this, a large-scale mesh of networked chips are required below the metasurface, which apart from adapting electrical impedance properties, also communicate with each other, thus relaying information about meta-atom settings, as well as forwarding possible distributed measurements taken. This paper describes why an asynchronous mixed-signal ASIC is advantageous for the control of scalable, EM absorbing, metasurfaces.  

\end{abstract}

\IEEEpeerreviewmaketitle
\IEEEoverridecommandlockouts

\section{Introduction}

\IEEEPARstart{M}{etamaterials} are artificial structures engineered to exhibit EM properties not commonly found in nature. Permittivity ($\varepsilon$) and permeability ($\mu$) in naturally occurring materials such as glass or water are mostly positive. However, some metals such as silver and gold have negative permittivity at short wavelengths whilst materials such as a surface plasmon have either $\varepsilon$ or $\mu$ negative. What is important though is that there are no materials having both $\varepsilon$ and $\mu$ negative. Such metamaterial structures, can be fabricated, to have negative $\varepsilon$ and negative $\mu$, something that Veselago\cite{veselago} predicted back in 1968. In 2001, Smith et al. \cite{smith_2001} with the first experiments in microwaves, showed a pass band in various samples, where both $\varepsilon$ and $\mu$ appeared to be negative, thus also giving a negative refractive index.    
   Since then, these structures have shown extraordinary EM properties, and numerous novel devices have been reported in the literature e.g. invisibility cloaks \cite{invisible_cloak}, leaky-wave antennas \cite{leaky_wave}, super-lenses \cite{superlens}, anomalous reflectors \cite{anomalousreflector1}\cite{anomalousreflector2} and perfect absorbers \cite{absorber}, all demonstrating unnatural functionalities. Most of these metamaterial/metasurface structures have a fixed function configured for a very narrow frequency and angle of incidence, with no means of reconfiguring properties once fabricated. 
   
One attempt to make these materials reconfigurable used mechanical elements, which when placed under stress, change shape \cite{mechanical}, however this approach has very limited reconfigurability. More recently electrically reconfigurable, programmable \cite{reconfigurable} \cite{programmable} metamaterials have emerged, where a digital signal is used to change the voltage bias of a diode-based capacitor, hence giving two different capacitive states to the metamaterial atom (meta-atom). Although a major step forward, this solution still has major limitations in tuneability, given only two possible states for the meta-atom. 

This paper advocates the use of asynchronous digital circuits in the development of an application specific integrated circuit (ASIC), to be embedded within metasurface structures, in order to give multiple degrees of freedom in the adaptability of the meta-atoms. These ASICs must also include mini-routers to move information through an arbitarily-large meta-atom array, so as to be able to configure the complex load impedances at the different locations of the metasurface. Section II describes the system structure and constraint requirements in which the ASICs need to operate within, whilst Section III describes how the asynchronous communication scheme is better suited to address these challenges and why it is preferred over its synchronous counterpart.

\section{System Structure}

\subsection{Overview}

\begin{figure}[h]
\centering
\includegraphics[width=9cm, height= 4cm]{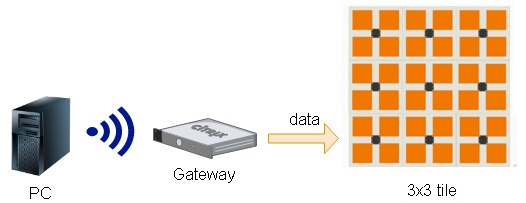}
\captionsetup{justification=centering}
\caption{Top Level System Architecture. After computing the metasurface configuration the computer sends the instructions wirelessly to the gateway that is physically located on the edge of the metasurface, which in turn sends instructions to the ASICs to configure their local meta-atom.}
\label{system_architecture}
\end{figure}

The system structure of a programmable metasurface, as shown in Figure \ref{system_architecture}, requires at least a 3 layer PCB, where the top-layer consists of an array of metal patches, e.g. as in \cite{vias}, the intermediate layer consists of a ground plane and the bottom layer consists of an array of ASICS, connected through vias to the top metal patch layer. The top layer is patterned according to the EM function to be implemented; options include anomalous reflection or absorbance. The intermediate layer is a ground plane so as to minimise the effect of the surface that metasurface conforms to. Finally, the ASICs role is to adapt the EM properties of the top-layer by providing adjustable complex-impedance loading, as well as networking functionality. The networking is required in order to receive and relay incoming commands. To keep the metasurface ASICs as simple as possible, a shared gateway connects the driving computer to the array. Figure 1 shows a possible configuration where an array of $3\times3$ ASICs (on the bottom layer) form a tile that consists of $6\times6$ metal patches on the top layer. After computing the metasurface configuration, based on incoming frequencies and angle of incidence information, the computer sends configuration settings, as desired complex impedance information for each meta-atom, to the tile via a wireless gateway. This gateway also supplies power to the metasurface ASICs via wired connections on the bottom plane. The tiles are such that an arbitrary number tiles can be clipped together to increase the size and shape of the overall metasurface.

\subsection{Requirements/Constraints}

The first metasurface prototype targeted by this work involves an adaptive absorber/anomalous reflector, for various angles of incidence. Its use might be for reducing interference of wifi signals, reducing human exposure to stray wifi signals over health concerns or even to intelligently reflect a wifi signal in a particular direction, so as to boost reception. In order to achieve this functionality, one needs adaptability for both active (resistive) and reactive (capacitive/inductive) components linking adjacent patches. To program the active and reactive components, a mixed-signal ASIC containing digitally controlled varactors and varistors is required. Furthermore, this ASIC must be able to communicate digitally with its adjacent neighbours in order to be able to pass on commands that are not destined for itself or in order to send messages to the gateway that concern a local measurement or event, such as a failure. Thus, key architectural choices must be made for these ASICs, taking into account the various requirements/constraints listed below:
 
\begin{itemize}

\item \textbf{Meta-atom Size}

Today wifi signals' frequency range varies between 2.4GHz (IEEE 802.11b) and 60GHz (IEEE 802.11ad), thus ranging in wavelength from $12.5cm$ to $5 mm$ respectively. In general, there should be \textit{at least} 5 meta-atoms per wavelength for correct operation and the tile length should include \textit{at least} $5\times5$ wavelengths for reasonable two-dimensional directivity. Thus scaling up the frequency of operation tightens the size constraints for the ASICs, given that the wavelength becomes smaller and the meta-atom needs to decrease in size; this is below $1mm\times1mm$ in the IEEE 802.11ad case. In this limited footprint, the meta-atom needs to accommodate the metal patches, all necessary components to tune the complex impedances between the patches, the control and communication circuitry, the pins and tracks to neighbouring meta-atoms and any other ``packaging inefficiencies". This makes the use of multiple components per meta-atom extremely challenging and thus directs towards a single chip ASIC for each meta-atom on the PCB.

\item \textbf{Conformal Metasurface}

In order to enable the metasurface to be applied to surfaces other than walls, e.g. on a vehicle, it is desired that the adaptive meta-atoms are designed on a flexible or rigid-flex PCB substrate.

\item \textbf{Scalability}

The tile must be easily scalable. By adding or removing metasurface patches, it should automatically be adjusted to recognize the number of patches in the tile and scale the addressing accordingly.
\\

\item \textbf{Low Cost}

The cost of a tile, and consequently the whole metasurface, must be low enough to be adopted. There is no point in engineering adaptive metasurfaces that can not be used due to prohibitive costs. This translates to as few components are necessary and to using a semiconductor technology that is not exotic and thus expensive. Furthermore the cost can be reduced if a single design of meta-atom ASIC, can cater for all cases in the array e.g. not have a different ASIC for the edges of the metasurface structure or for interfacing to the gateway.

\item \textbf{Power consumption}

The need for low power consumption is a fundamental limitation for most of the integrated circuits, but this is especially pertinent given the large surfaces to be covered and where the power can scale up very quickly. The power needs to be supplied by the gateway or if possible through energy scavenging techniques. 

\item \textbf{EM emissions}

The circuitry that enables the adaptability of the absorber/anomalous reflector should have minimal or if possible zero EM emissions so as to avoid interfering with the incident wave through reflections off the covered surface. A large surface that is clocked could end up radiating a large amount of energy into an area that was supposed to eliminate existing EM signals, and hence exasperate the problem that was to be solved. 

\item \textbf{Timing Constraints}  

In a synchronous ASIC digital solution, part of the design methodology requires the synthesis of a clock tree that is scalable. It should also satisfy the requirements of a flexible metasurface where tiles are connected together to create a surface as large as a wall and possibly conforming to irregular shapes. The clock tree should be resilient to process variability as well as variability in the overall metasurface. Thus it should provide margins for both intra and inter die communications. Furthermore, the clock skew needs to be well controlled in order to prevent the violation of the setup and hold time constraints. Given that the same dies should be ``generic" in the sense that they may be used to populate different metasurface designs this is not trivial, if not impossible. Today's design tools are optimised for synthesizing clock trees on a single chip and at most to the area of a motherboard. Furthermore, it has been reported \cite{power_clock_tree} that in modern chips, in order to conform to timing constraints, an elaborate clock tree is synthesised such that as much as $40\%$ of the total power consumption is utilized by just the clock-tree. An alternate solution would be to use a crystal oscillator in each meta-atom, and use a phase-locked loop to bring all clocks in sync. However, since crystals are relatively expensive and their sizes are at least a few mm in size this is not a viable solution.

\end{itemize}

The above constraints make an asynchronous-digital based, mixed-signal ASIC an attractive solution. Though an approach that is not used often, due to the design complexity and the lack of the commercially available tools for automated synthesis, for this application asynchronous design seems to be the most suitable choice. The next section describes the characteristics of asynchronous circuits and why they are an enabling technology for this application.
    
\section{Asynchronous vs Synchronous for Adaptive Metasurface ASICs}

The meta-atoms' ASICs require a mixed-signal solution where a number of D/A converters are used to control varactors and varistors inserted in the RF signal paths. The choice of the type of digital design is a key aspect that defines the viability of a highly adaptive metasurface. Digital circuit design can be divided into two opposing methodologies, that concern timing constraints. The most dominant approach is that of synchronous circuits, whereby signals synchronise changes of states across the entire chip, at discrete points in time. The global clock is therefore a really powerful technique for simplifying the implementation of sequencing circuits. However, as the size of a system increases, the common and discrete notion of time becomes more difficult to achieve, especially when the size of the system can vary dynamically. Asynchronous circuits are different. There is no common and discrete time. For the communication between components, handshaking is used to coordinate the movement of information and computation. In the following section a brief introduction to the basic asynchronous circuits is presented and the suitability, as opposed to a synchronous design, for large adaptive metasurfaces is discussed.

\subsection{Introduction to Asynchronous Circuits and Handshaking}

The principle behind asynchronous logic design is that each block or subsystem only communicates with adjacent blocks when the exchange of information information is desired. The communication does not have to wait to be triggered at every clock cycle but can occur at arbitrary times. The key element making this possible is the \textit{handshake}. 

\begin{figure}[h]
\centering
\includegraphics[width=85mm, height= 65mm]{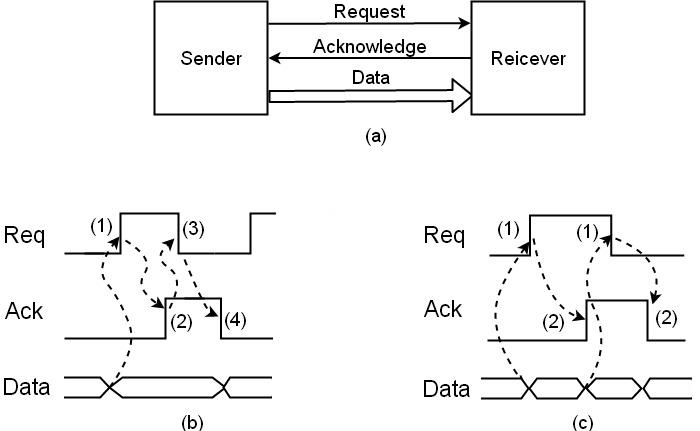}
\captionsetup{justification=centering}
\caption{(a) The communication scheme between the sender and the receiver using the request and acknowledge signals, (b) the four-phase protocol signal transition for one event and (c) the two-phase protocol signal transition for one event.}
\label{bundle_data}
\end{figure}

A variety of handshake protocols have been reported in the literature \cite{sparso}. The most common being the four-phase and the two-phase bundled data protocols. Both protocols utilize the request and acknowledge signals, along with a data line or a data bus as shown in Figure \ref{bundle_data}a. The four-phase communication protocol, illustrated in Figure \ref{bundle_data}b goes as follows: (1) once the data is made available on the data line, the sender (left) sets the request HIGH and (2) when ready, the receiver latches the data and sets the acknowledge HIGH, (3) the sender then takes request LOW and (4) the receiver retracts the acknowledge signal going back LOW, opening the way for a new communication to begin. The four-phase bundled data protocol has the advantage that the handshake circuits are very simple, but has the disadvantage of using unnecessary time and power for lowering the two signals. This could be avoided by using the two-phase bundled data protocol (Figure \ref{bundle_data}c) since with this protocol, it is each signalling transition 0$\rightarrow$1 and 1$\rightarrow$0 that represents the event, rather than the actual level on the request and acknowledge lines. Two-phase signalling requires slightly more complex circuits to be implemented. There are other protocols, e.g. single rail encoding, dual rail encoding, etc, with other pros and cons, however these will not be elaborated here.  

At the circuit level, an important building block is the Muller-C element, which in its basic form changes its output only when all the inputs ``agree" i.e. if all inputs are zero then the output transitions to zero, if all inputs are 1 then the output transitions to 1, for anything else, the state of the element remains the same. It is very useful for creating a signal to indicate when all parallel threads have been completed, in addition to implementing more simple handshakes.

\begin{figure}[h]
\centering
\includegraphics[width=90mm, height= 65mm]{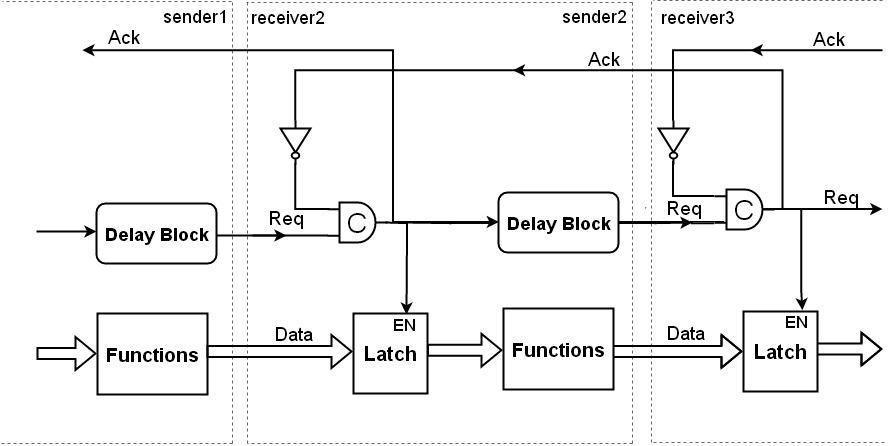}
\captionsetup{justification=centering}
\caption{The four-phase bundle-data implementation. The request signal is being delayed until the function block is finished so as to keep synchronization between the data and the handshake signals.}
\label{four_phase_implementation}
\end{figure}

Figure \ref{four_phase_implementation} shows the implementation of the handshake circuit between two modules. The function blocks start their operation once the request signal has been received. The delay block is added to prevent enabling(request signal high) the latch before the data are ready to be stored. Then the Muller-C element is enabled and the data are stored and move to the next block (if needed). Once the Muller-C element goes HIGH, the acknowledge signal is sent back to the previous controller to inform that the data have been received.

A variety of asyncronous circuits have been presented over the years exploiting the circuit level handshake communications and enabling data driven asynchronous circuit design. Fast and low power microprocessors \cite{asynchronous_microprocessor_2}, 3D image processors \cite{imager}, filter banks for digital hearing aid \cite{filter} and wireless endoscopic capsule \cite{capsule} are some of the applications. Their characteristics are known but they have not been adopted for mainstream applications. 

\subsection{The Case for Asynchronous ASICs in Metasurfaces}

Although synchronous digital is by far the most predominant methodology used today, there are special cases where the inconvenience of not having commercial CAD tools and readily available libraries to automate the design process, is still not an issue. In the case of an adaptive metasurface ASIC, the difficulty of producing a reliable clock signal over such a large surface makes the asynchronous route look attractive since asynchronous signalling \cite{asynchronous_microprocessor_2} \textit{can easily cope with added parasitics or extended/flexed external communication lines needed on a conformal metasurface}, as long as these affect the data and handshaking lines equally. This property stems from the circuits inherent delay-insensitivity \cite{scalability}\cite{supply_noise}. In such a case the handshake process may be slowed down, however it will conclude successfully. This argument also applies to easy scalability, since \textit{composing asynchronous systems is simply a matter of connecting the proper modules with matching interfacing specifications} \cite{interface}. This leads to ease of incremental improvement, better technology migration and be able to reuse most of the modules. Furthermore, an asynchronous solution means that there is \textit{no reason to include crystal based oscillators} within the dense array, thus saving space and reducing the cost. Typically a crystal oscillator measures a few $mm^2$ in size that is comparable to size of the ASIC. 

Asynchronous circuits are considered to be \textit{extremely energy-efficient} since they only charge and discharge capacitive nodes when they have to. Systems driven by a clock consume power every time there is a clock event even if no data has changed thus wasting energy. It is possible to have multiple clock domains and to disable parts of the clock tree, however such methods need elaborate control schemes to be driven at a high level. As an example of low-power application, Paver \cite{async_micro} presented an asynchronous version of a microprocessor, which is one of the most energy-efficient microprocessors compared to both synchronous and asynchronous schemes. If an older or low-power semiconductor technology with higher thresholds is chosen then device leakage can be considered negligible and so when no settings in the metasurface are being changed the power consumption is negligible. 

A recent work in the University of Utah \cite{async_micro_vs_sync_micro} compares the same microprocessor first by using a clocked network and then by using a control path. The results show a 5\% smaller area for the asynchronous design (control path) but with a slightly higher leakage energy due to more buffers that have been used. What is more important though is the 10 times average reduction in power. At a time where the power dissipation is the most significant limitation in circuits, being able to achieve 10 times less power dissipation is a huge step forward in the digital world.

\textit{Electromagnetic emissions in asynchronous circuits have been shown to be less} than the synchronous counterparts \cite{less_emissions} whilst also giving more \textit{evenly spread emissions} \cite{evenly_spread_emissions}. This is expected since by not synchronising all switching activity the signals can not all add up by being in phase. Asynchronous circuits are considered \textit{faster} \cite{speed} compared to the synchronous because they exhibit \textit{average-case behaviour} instead of being driven by a clock that assumes worst-case behaviour for all blocks. They also have completion detection mechanisms to detect when an operation is finished, which means the next operation can start.
Digital asynchronous circuits have the added advantage of generating less power supply noise \cite{supply_noise}, when compared to their synchronous counterpart given that the current draw is better distributed in time.

\section{Conclusion}

In this paper, we have presented a targeted architecture of a large, adaptive metasurface that can both absorb or anomalously reflect EM waves, that has particularly challenging requirements and physical constraints that make designing with a synchronous methodology very difficult. As a result, we introduced and explored the properties of asynchronous digital circuits as an enabling technology for programmable metamaterials, with clear benefits with regards to doing away with the global clock to achieve delay insensitivity, low EM emissions and low power, amongst other benefits. 

\section{Acknowledgements}
This work is partially funded from the European Union's Horizon 2020 Programme FETOPEN-2016-2017, VISORSURF project, under grant agreement no. 736876.

\bibliography{IEEEabrv,references}

\begin{thebibliography}{10}
\providecommand{\url}[1]{#1}
\csname url@samestyle\endcsname
\providecommand{\newblock}{\relax}
\providecommand{\bibinfo}[2]{#2}
\providecommand{\BIBentrySTDinterwordspacing}{\spaceskip=0pt\relax}
\providecommand{\BIBentryALTinterwordstretchfactor}{4}
\providecommand{\BIBentryALTinterwordspacing}{\spaceskip=\fontdimen2\font plus
\BIBentryALTinterwordstretchfactor\fontdimen3\font minus
  \fontdimen4\font\relax}
\providecommand{\BIBforeignlanguage}[2]{{%
\expandafter\ifx\csname l@#1\endcsname\relax
\typeout{** WARNING: IEEEtran.bst: No hyphenation pattern has been}%
\typeout{** loaded for the language `#1'. Using the pattern for}%
\typeout{** the default language instead.}%
\else
\language=\csname l@#1\endcsname
\fi
#2}}
\providecommand{\BIBdecl}{\relax}
\BIBdecl

\bibitem{veselago}
V.~G. Veselago, ``The electrodynamics of substances with simultaneously
  negative values of {$\epsilon$} and μ,'' \emph{Soviet Physics Uspekhi},
  vol.~10, no.~4, p. 509, 1968.

\bibitem{smith_2001}
R.~A. Shelby, D.~R. Smith, and S.~Schultz, ``Experimental verification of a
  negative index of refraction,'' \emph{Science}, vol. 292, no. 5514, pp.
  77--79, 2001.

\bibitem{invisible_cloak}
D.~Schurig, J.~J. Mock, B.~J. Justice, S.~A. Cummer, J.~B. Pendry, A.~F. Starr,
  and D.~R. Smith, ``Metamaterial electromagnetic cloak at microwave
  frequencies,'' \emph{Science}, vol. 314, no. 5801, pp. 977--980, 2006.

\bibitem{leaky_wave}
K.~M. Kossifos and M.~A. Antoniades, ``A zero beam-squinting leaky-wave antenna
  using nri-tl metamaterials,'' in \emph{2015 9th European Conference on
  Antennas and Propagation (EuCAP)}, May 2015, pp. 1--2.

\bibitem{superlens}
W.~Cai, D.~A. Genov, and V.~M. Shalaev, ``Superlens based on metal-dielectric
  composites,'' \emph{Phys. Rev. B}, vol.~72, p. 193101, Nov 2005.

\bibitem{anomalousreflector1}
A.~Díaz-Rubio, V.~Asadchy, A.~Elsakka, and S.~Tretyakov, ``Metasurfaces for
  perfect control of reflection,'' in \emph{2017 International Workshop on
  Antenna Technology: Small Antennas, Innovative Structures, and Applications
  (iWAT)}, March 2017, pp. 3--5.

\bibitem{anomalousreflector2}
V.~S. Asadchy, A.~D\'{\i}az-Rubio, S.~N. Tcvetkova, D.-H. Kwon, A.~Elsakka,
  M.~Albooyeh, and S.~A. Tretyakov, ``Flat engineered multichannel
  reflectors,'' \emph{Phys. Rev. X}, vol.~7, p. 031046, Sep 2017.

\bibitem{absorber}
D.~Zhirihin, C.~Simovski, P.~Belov, and S.~Glybovski, ``Mushroom high-impedance
  metasurfaces for perfect absorption at two angles of incidence,'' \emph{IEEE
  Antennas and Wireless Propagation Letters}, vol.~16, pp. 2626--2629, 2017.

\bibitem{mechanical}
B.~Florijn, C.~Coulais, and M.~van Hecke, ``Programmable mechanical
  metamaterials,'' \emph{Phys. Rev. Lett.}, vol. 113, p. 175503, Oct 2014.

\bibitem{reconfigurable}
T.~Jun~Cui, M.~Q. Qi, X.~Wan, J.~Zhao, and Q.~Cheng, ``Coding metamaterials,
  digital metamaterials and programmable metamaterials,'' vol.~3, p. e218, 10
  2014.

\bibitem{programmable}
H.~Yang, X.~Cao, F.~Yang, J.~Gao, S.~Xu, M.~Li, X.~Chen, Y.~Zhao, Y.~Zheng, and
  L.~Sijia, ``A programmable metasurface with dynamic polarization, scattering
  and focusing control,'' vol.~6, 10 2016.

\bibitem{vias}
O.~Luukkonen, F.~Costa, C.~R. Simovski, A.~Monorchio, and S.~A. Tretyakov, ``A
  thin electromagnetic absorber for wide incidence angles and both
  polarizations,'' \emph{IEEE Transactions on Antennas and Propagation},
  vol.~57, no.~10, pp. 3119--3125, Oct 2009.

\bibitem{power_clock_tree}
A.~Kapoor, C.~Groot, G.~V. Piqué, H.~Fatemi, J.~Echeverri, L.~Sevat,
  M.~Vertregt, M.~Meijer, V.~Sharma, Y.~Pu, and J.~P. de~Gyvez, ``Digital
  systems power management for high performance mixed signal platforms,''
  \emph{IEEE Transactions on Circuits and Systems I: Regular Papers}, vol.~61,
  no.~4, pp. 961--975, April 2014.

\bibitem{sparso}
J.~Spars{\o}, ``Asynchronous circuit design - a tutorial,'' in \emph{Chapters
  {1-}8 in {''}Principles of asynchronous circuit design - A systems
  Perspective''}.\hskip 1em plus 0.5em minus 0.4em\relax Boston / Dordrecht /
  London: Kluwer Academic Publishers, dec 2001, pp. 1--152.

\bibitem{asynchronous_microprocessor_2}
A.~Takamura, M.~Kuwako, M.~Imai, T.~Fujii, M.~Ozawa, I.~Fukasaku, Y.~Ueno, and
  T.~Nanya, ``Titac-2: an asynchronous 32-bit microprocessor based on
  scalable-delay-insensitive model,'' in \emph{Proceedings International
  Conference on Computer Design VLSI in Computers and Processors}, Oct 1997,
  pp. 288--294.

\bibitem{imager}
J.~Georgiou, A.~G. Andreou, and P.~O. Pouliquen, ``A mixed analog/digital
  asynchronous processor for cortical computations in 3d soi-cmos,'' in
  \emph{2006 IEEE International Symposium on Circuits and Systems}, May 2006,
  pp. 4 pp.--.

\bibitem{filter}
L.~S. Nielsen and J.~Sparso, ``Designing asynchronous circuits for low power:
  an ifir filter bank for a digital hearing aid,'' \emph{Proceedings of the
  IEEE}, vol.~87, no.~2, pp. 268--281, Feb 1999.

\bibitem{capsule}
X.~Zhang, H.~Jiang, and Z.~Wang, ``Using asynchronous circuits for
  communications in wireless endoscopic capsule,'' in \emph{APCCAS 2008 - 2008
  IEEE Asia Pacific Conference on Circuits and Systems}, Nov 2008, pp.
  1244--1247.

\bibitem{scalability}
A.~J. Martin, ``Compiling communicating processes into delay-insensitive vlsi
  circuits,'' \emph{Distributed Computing}, vol.~1, no.~4, pp. 226--234, Dec
  1986.

\bibitem{supply_noise}
C.~D. Nielsen and A.~J. Martin, ``Design of a delay-insensitive
  multiply-accumulate unit,'' \emph{Integration, the VLSI Journal}, vol.~15,
  no.~3, pp. 291 -- 311, 1993, special Issue on asynchronous systems.

\bibitem{interface}
C.~Niessen, C.~H. van Berkel, M.~Rem, and R.~W. J.~J. Saeijs, ``Vlsi
  programming and silicon compilation-a novel approach from philips research,''
  in \emph{Proceedings 1988 IEEE International Conference on Computer Design:
  VLSI}, Oct 1988, pp. 150--151.

\bibitem{async_micro}
N.~C. Paver, ``The design and implementation of an asynchronous
  microprocessor,'' Ph.D. dissertation, Department of Computer Science,
  University of Manchester, 1994.

\bibitem{async_micro_vs_sync_micro}
D.~Bhadra and K.~S. Stevens, ``Design of a low power, relative timing based
  asynchronous msp430 microprocessor,'' in \emph{Proceedings of the Conference
  on Design, Automation \& Test in Europe}, ser. DATE '17.\hskip 1em plus 0.5em
  minus 0.4em\relax 3001 Leuven, Belgium, Belgium: European Design and
  Automation Association, 2017, pp. 794--799.

\bibitem{less_emissions}
N.~C. Paver, P.~Day, C.~Farnsworth, D.~L. Jackson, W.~A. Lien, and J.~Liu, ``A
  low-power, low noise, configurable self-timed dsp,'' in \emph{Proceedings
  Fourth International Symposium on Advanced Research in Asynchronous Circuits
  and Systems}, Mar 1998, pp. 32--42.

\bibitem{evenly_spread_emissions}
C.~Myers, \emph{Asynchronous Circuit Design}.\hskip 1em plus 0.5em minus
  0.4em\relax Wiley, 2004.

\bibitem{speed}
A.~J. Martin, A.~Lines, R.~Manohar, M.~Nystrom, P.~Penzes, R.~Southworth,
  U.~Cummings, and T.~K. Lee, ``The design of an asynchronous mips r3000
  microprocessor,'' in \emph{Proceedings Seventeenth Conference on Advanced
  Research in VLSI}, Sep 1997, pp. 164--181.

\end{thebibliography}
\bibliographystyle{IEEEtran}

\end{document}